\documentclass[twocolumn,aps,pre,10pt]{revtex4-1}

\usepackage{color}
\usepackage{subfig}
\usepackage{graphicx}
\usepackage{hyperref}

\begin{document}

\title{Predicting Efficiency in master-slave grid computing systems}
\author{Gonzalo Travieso}
\author{Carlos A. Ruggiero}
\author{Odemir M. Bruno}
\author{Luciano da F. Costa}
\affiliation{Instituto de F\'isica de S\~ao Carlos, Universidade de
  S\~ao Paulo, Brazil}
\date{\today}

\begin{abstract}
  This work reports a quantitative analysis to predicting the
  efficiency of distributed computing running in three models of
  complex networks: Barabási-Albert, Erd\H{o}s-Rényi and
  Watts-Strogatz.  A master/slave computing model is simulated.  A
  node is selected as master and distributes tasks among the other
  nodes (the clients).  Topological measurements associated with the
  master node (e.g.\ its degree or betwenness centrality) are
  extracted and considered as predictors of the total execution time.
  It is found that the closeness centrality provides the best
  alternative. The effect of network size was also investigated.
\end{abstract}

\maketitle

\section{Introduction}
\label{sec:intro}

Despite all the scientific and technological advances in scientific
computing obtained along the last decades, the demand on intensive
number crunching remains as high as ever.  This has been caused by at
least the three following factors: (i)~increasing amount of data being
continuously produced in virtually all areas; (ii)~the higher
resolutions of such data, often involving 3D images and movies; and
(iii)~the focus on complex, non-linear systems that has underlain much
of the basic and applied sciences.  Given that the individual
processing power has saturated~\cite{geer05}, ever increasing
attention has been placed on distributed and networked systems.  One
particularly attractive possibility, known as \emph{grid-computing}
involves the use of a large number of computers, duly connected to the
Internet, as a massive computing resource~\cite{foster03:_grid}.  The
immediate advantage of such an approach is to provide the access to a
large mass of already existent machines, without much additional cost
for hardware or software.  However, the performance of a grid system
will depend strongly on the protocol for distributing the tasks, the
topology of the interconnections, and the type of processing required.
As a matter of fact, the execution of distributed tasks in a grid
network is a largely complex problem directly related to the
structure/function paradigm in complex networks~\cite{boccaletti06}.
Previous approaches at trying to understand the efficiency in grid
computing include a study of the effect of network connectivity
\cite{complexgrid} and of geographical constraints \cite{geographical}
on a simple grid computing model and how topology affects applications
with real-time constraints \cite{realtime}.  Despite such recent
advances, the investigation of how the topology of the
interconnections on the respective performance remains an open
problem.  In particularly, it would be interesting to devise means to
predict the execution time of grid computing systems from simpler
network measurements, without the need of undergoing the whole process
of real computations.  The present work aims precisely at
investigating how several types of networks
measurements~\cite{measure} can be used to foresee the execution time
in grid computing systems as modeled and simulated in terms of model
complex networks of varying topology.

\section{Methodology}
\label{sec:meth}

\subsection{Grid Representation as a Complex Network}

In this work, computing grids are modeled as graphs where the nodes
are computers and the edges are two-directional internet-like
connections (e.g.\ TCP/IP). This gives rise to an undirected complex
network with $N$ nodes, that can be represented by a respective
\emph{adjacency matrix}, A. The element of the $i$-th row and $j$-th
column of the matrix, denoted by $A_{ij}$, has value~1 if and only if
there is a link between the $i$-th and the $j$-th node, and the value
0, otherwise.

The main issue in the current work is to choose a computer that will
distribute the tasks to all other computers in such a way as to
minimize the total execution time. It is natural to suppose that this
optimum node should be, in some sense, \emph{central} to the grid
topology. With this in mind, the following metrics have been chosen in
order to characterize the centrality of each node: \emph{degree
  centrality}, \emph{betweenness centrality} and \emph{closeness
  centrality} (related with the \emph{average distance}). Another
important measure was included: the local clustering coefficient.  All
these metrics can be derived from the adjacency matrix $A$.

Consider a node $i$.  The degree centrality (also known as \emph{node
  degree}) gives the number of adjacent links to $i$ and is
easily calculated by taking the number of non-zero elements in row
$i$:
$$ \sum_{j} A_{ij}
$$
The betweenness centrality tries to quantify how much a node is in the
path connecting two other nodes. Considering all shortest paths
connecting all pairs of nodes, the betweenness centrality of $i$ is
related to the number of these paths passing through $i$ and defined
as
$$ \sum_{j, k \ne i} \frac{s_{jk}(i)}{s_{jk}}
$$
where $s_{ij}$ is the total number of shortest paths between nodes $j$
and $k$, and $s_{jk}(i)$ is the number of those paths that pass
through node $i$.  In this work we used the algorithm of Brandes
\cite{brandes01} for its computation.  The closeness centrality of
node $i$ quantifies how close it is to the other nodes in the network
and is taken as the reciprocal of the mean of the distances from $i$
to all other nodes:
$$ \frac{N}{\sum_j d_{ij}}
$$
where $d_{ij}$ is the distance between nodes $i$ and $j$ and is taken
as the number of edges of the shortest path connecting them.  Finally,
the local clustering coefficient of $i$ is defined as the number of
links connecting its neighbors divided by the number of links
necessary to fully connect all those neighbors (the number of links
necessary to connect each and every pair of neighboring nodes),
leading to a number between 0 and 1:
$$ \frac{e_i}{\frac{k_i(k_i-1)}{2}}
$$
where $e_i$ is the number of links among neighbors of node $i$.

\subsection{Network Models}

Three complex network models were chosen for this work:
Erd\H{o}s-R\'{e}nyi (ER), Barabási-Albert (BA) and
Watts-Strogatz (WS) \cite{newman03}.

In the ER model used fixes the number of nodes in the network $N$ and
the probability $p$ of creating a link between any two nodes.

The Erd\H{o}s-R\'{e}nyi model is interesting for comparison and
analysis but it lacks some properties found in real computing
networks.  Preferential attachment models like the Bar\'{a}basi-Albert
(BA) are more realistic, because they generate networks with degree
distributions following a power law, very similar to those found in
the real Internet \cite{faloutsos99}. Considering that many grid
computing infrastructures are built upon the Internet or upon an
Internet like network, the BA model is included in our analysis. BA
networks are built by starting with a few unconnected nodes; for each
new node created, $m$ connections are generated from this node to
others in such a way that nodes with higher degree are more likely to
receive one of the $m$ connections.  The resulting network average
degree will be equal to $2m$.

As another example of a network model that has different properties
than the aforementioned, we chose the Watts-Strogatz (WS) model which
implements the ``small-world effect.'' Short distances are frequently
desirable in grids in order to minimize the communication latency
among nodes. This model combines clustering coefficients that are
higher than those in the BA and ER models with small distances. It
starts with a regular structure such as a ring, with average degree
$k$ and then rewires some edges with uniform probablity $r$.  Larger
values of $r$ decrease clustering and average distances, but distances
fall much faster.  It is therefore possible, by adjusting $r$, to see
the effect of high clustering and small distances simultaneously.

\subsection{Computational model}
\label{sec:comput}

\begin{figure*}
  \begin{centering}
  \subfloat[Computational model]{\includegraphics[width=0.49\textwidth]{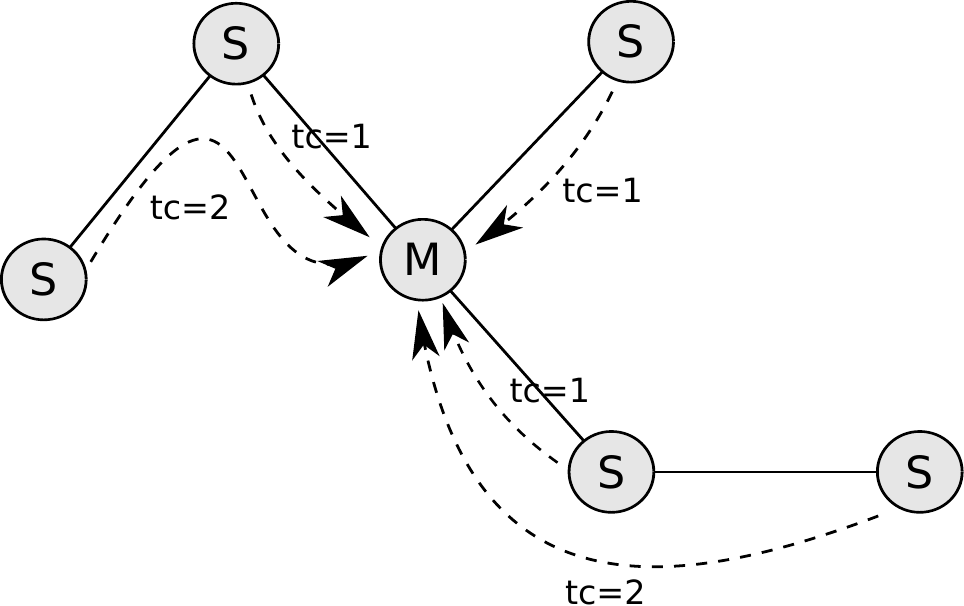}}
  \subfloat[Steps of task execution]{\hspace{0.11\textwidth}\includegraphics[width=0.27\textwidth]{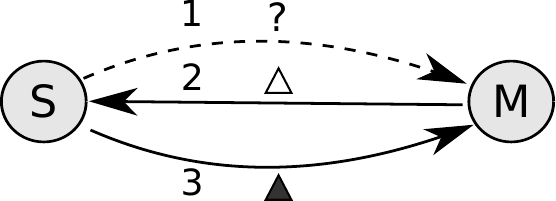}\hspace{0.11\textwidth}}
  \end{centering}
  \caption{(a) Diagram showing the computation model, each connection
    requires the same amount of time (one unit). (b) Steps of the
    model, 1)~slave sends a task request to the master, 2)~master
    sends a task to be computed and 3)~Slave returns the result to the
    master}
  \label{fig:computerModel}
\end{figure*}

The computational model used is the same introduced in
Ref.~\cite{complexgrid}.  The application works as master/slave, with
a master having a set of tasks that are distributed to the slaves for
processing.  All tasks require the same amount of time to compute.
After choosing one of the nodes as the master, all other nodes in the
network are slaves. Figure \ref{fig:computerModel} shows a diagram of
the computational model, which illustrates the time necessary for the
slaves to send a message to the master. The processing starts with the
slaves sending a task request to the master.  For each request
received, the master sends a task to be computed.  After the slave
finishes computing the task, the result is returned to the master,
which sends another task to the slave, if there are yet tasks to
compute (see Figure \ref{fig:computerModel}).  When the results of all
tasks are received by the master, the application stops.

We consider that the time to send a task or result through each link
is the same, and use this as the time unit.  We also assume that the
time taken by the master to choose a task to send and by each node to
route packets to a neighbor node are negligible. Also, all nodes have
the same computational power.

At the start of the execution, a slave that lies at a distance $d$
from the master will need to wait for a task for $2d$ time units ($d$
units for its request to reach the master and another $d$ for the task
to arrive).  Suppose the node takes $c$ time units to compute the
task.  The master will then receive the result for the first task from
this slave at $2d + c + d = 3d + c$.  Subsequent task results will
arrive from this slave at intervals of $2d + c$ (because the arrival
of a result is a sign that the slave is idle, and the master can send
a new task).  If the total number of tasks is sufficiently large, all
slaves at distance $d$ from the master will receive approximately the
same number of task to compute; we represent this number of tasks as
$m(d)$, as it depends on the distance.  They will therefore take a
total time of
\begin{equation}
  \label{eq:timed}
  T(d) = d + m(d) (2d + c).
\end{equation}

Again if the total number of tasks is sufficiently large, all slaves
will end processing approximately at the same time, such that we can
consider $T(d)$ in Equation~\ref{eq:timed} as the total execution
time, but we will further simplify the expression because in this case
$m(d) (2d + c)$ is large with respect to $d$, giving:
\begin{equation}
  \label{eq:time}
  T = m(d) (2d + c).
\end{equation}

If we have a total of $M$ tasks to execute we get:
\begin{equation}
  \label{eq:M}
  M = \sum_{d=1}^{d_\mathrm{max}}n(d)m(d) =
  \sum_{d=1}^{d_\mathrm{max}}\frac{n(d)T}{2d+c},
\end{equation}
where $n(d)$ is the number of nodes at distance $d$ from the master
and $d_\mathrm{max}$ is the maximum distance from the master for all
nodes.  Isolating $T$ in Equation~\ref{eq:M} we get
\begin{equation}
  \label{eq:T}
  T = \frac{M}{\sum_{d=1}^{d_\mathrm{max}}\frac{n(d)}{2d+c}}.
\end{equation}

If we have $c \gg 2d$, Equation~\ref{eq:T} reduces to
$$
  T = \frac{Mc}{N-1},
$$
where $N$ is the number of nodes in the network.  This is the best
possible result for this application, and is achieved independently
from network topology.  On the other hand, if $c \ll 2d$, most time is
taken sending and receiving tasks, and it will make no sense to use a
grid computing system (better results would be achieved by local
execution of the tasks).  The interesting values are then for $c
\approx 2d$, where total execution time depends on topology.  More
specifically, the results depend on the hierarchical structure of the
network \cite{hierarchical}.

\section{Results and discussion}
\label{sec:res}

In order to quantify the overall processing performance, we adopt the
concept of \emph{efficiency}~\cite{padua11}.  This measurement is
calculated dividing the achieved speed-up by the ideal speed-up.
Speed-up refers to the ratio between the total execution time with one
processor and the total execution time using all the processors in the
network.  Therefore, the ideal speed-up is $N$, i.e.\ the number of
available processors (nodes).

The performed experiments proceed in three stages.  First we compare
the potential of several network topology measurements in predicting
the grid efficiency for a BA configuration.  Then, we investigate the
effect of network size in the prediction.  Finally, we consider other
network models, more specifically the ER and WS models.

\subsection{Preliminary Analysis}

\begin{figure*}[htbp]
  \begin{centering}
  \subfloat[Efficiency $\times$ Average Degree]{\includegraphics[width=0.5\textwidth]{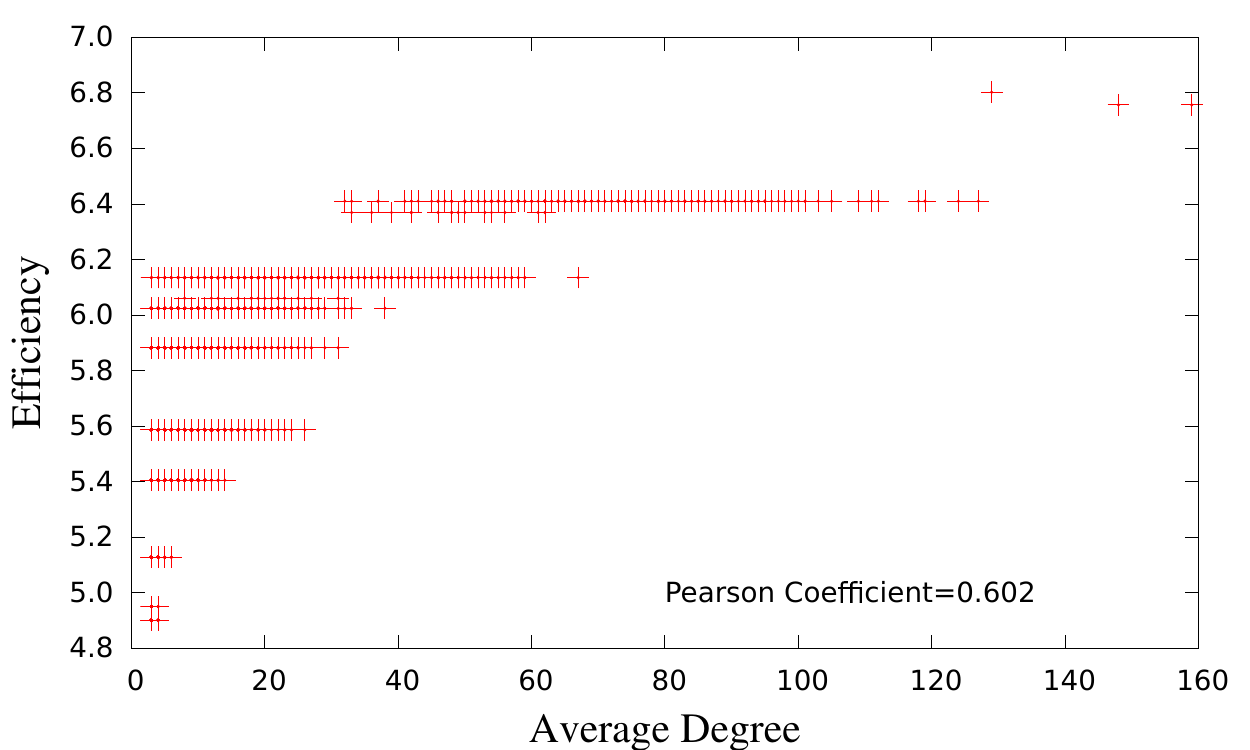}}
  \subfloat[Efficiency $\times$ Clustering Coefficient]{\includegraphics[width=0.5\textwidth]{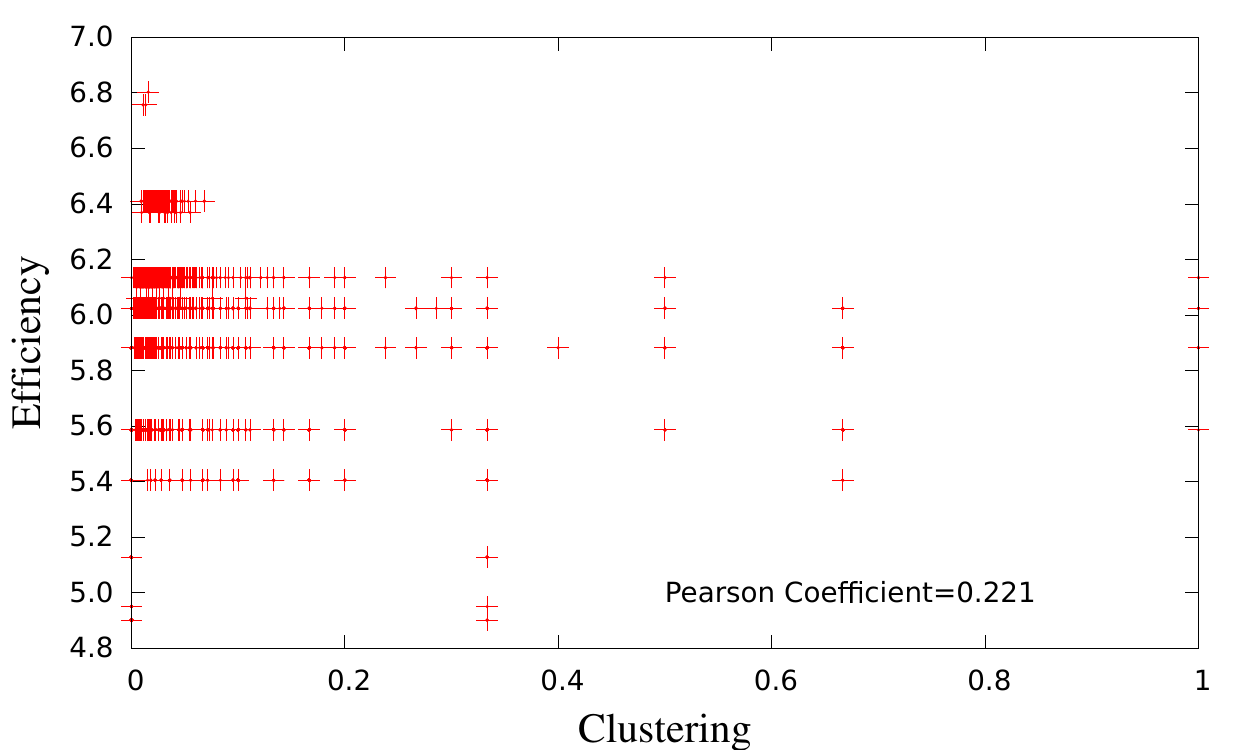}}\\
  \subfloat[Efficiency $\times$ Betweeness]{\includegraphics[width=0.5\textwidth]{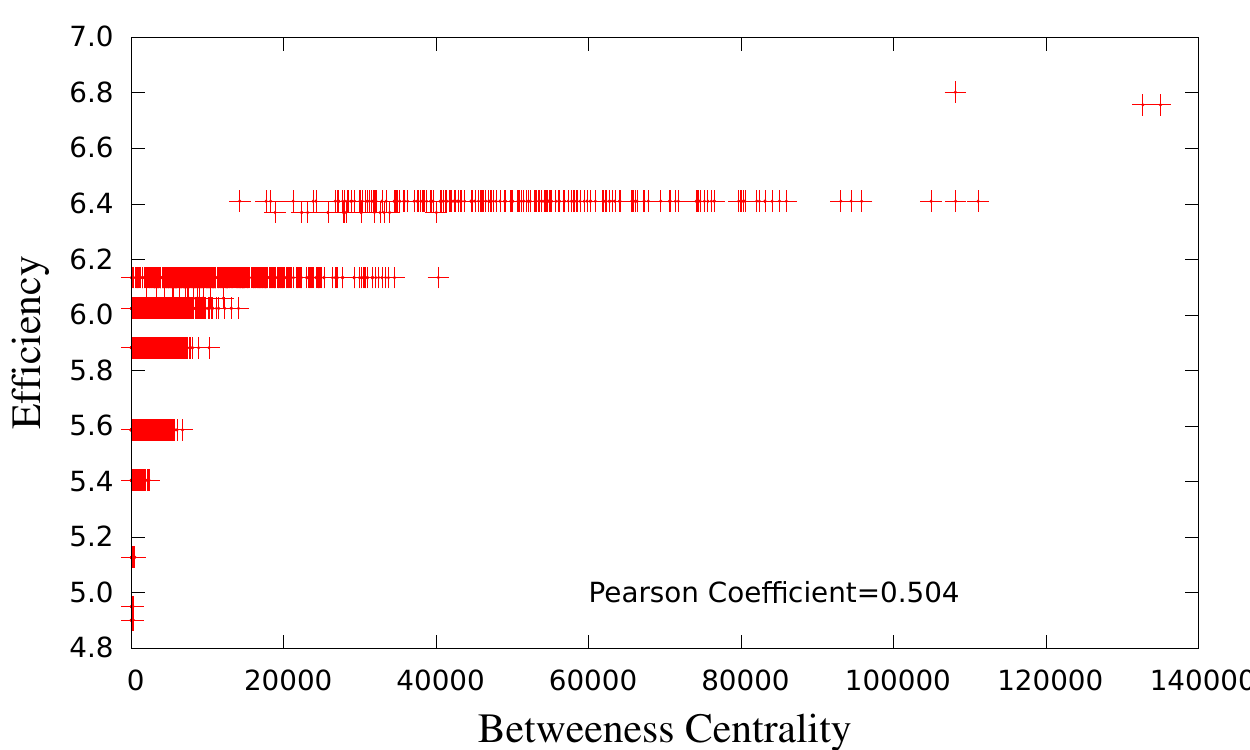}}
  \subfloat[Efficiency $\times$ Closeness]{\includegraphics[width=0.5\textwidth]{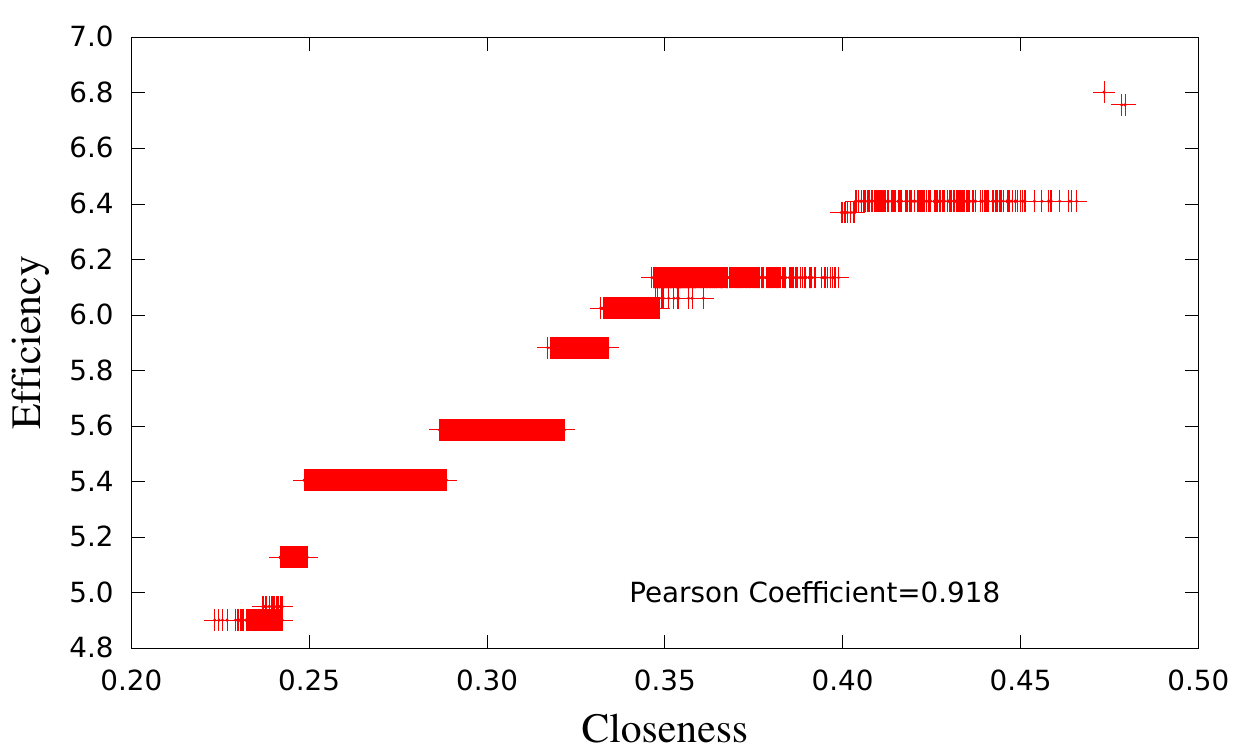}}
  \end{centering}
  \caption{Efficiency $\times$ Measure for Barabasi model with 1000 nodes and average degree 6}\label{effxmeas}
\end{figure*}

In order to have an initial indication of how several measurements of
the topology of the networks influence execution time, we considered a
BA networks with $N=1000$ and $\langle k \rangle> = 6$ (the latter
reflects the trend observed in the real internet~\cite{newman03}).
The individual tasks to be performed were assumed to be 10 (in the
choosen time unit that corresponds to the time to send a task through
a network link).  Each simulation involves 10000 tasks.  We obtained
the execution times for 30 network configurations, which had their
topology characterized in terms of their average degree, clustering
coefficient, betweeness and closeness centralities.
Figure~\ref{effxmeas} shows the processing efficiency in terms of each
of these measurements, with respective Pearson correlation
coefficients collected in Table~\ref{measure_pearson}.  It is clear
that the closeness exhibited a significant correlation with efficiency
and the largest Pearson coefficient when compared with the other
measurements, suggesting that it can be used to provide accurate
predictions of the processing efficiency.  For this reason, we
henceforth focus on the closeness in this work.

\begin{table}
  \caption{Average (30 samples) Pearson coefficient value }
  \label{measure_pearson}
  \begin{tabular}{lr}
    \toprule
    \multicolumn{1}{c}{Measure} & \multicolumn{1}{c}{Average Pearson} \\ 
    \colrule             
    Degree       & 0.611 \\
    Clustering   & 0.212 \\
    Closeness    & 0.918 \\
    Betweenness  & 0.514 \\
    \botrule
  \end{tabular}
\end{table}

\subsection{Effect of Network Size}

\begin{figure*}[htbp]
  \begin{centering}
  \subfloat[N=100]{\includegraphics[width=0.5\textwidth]{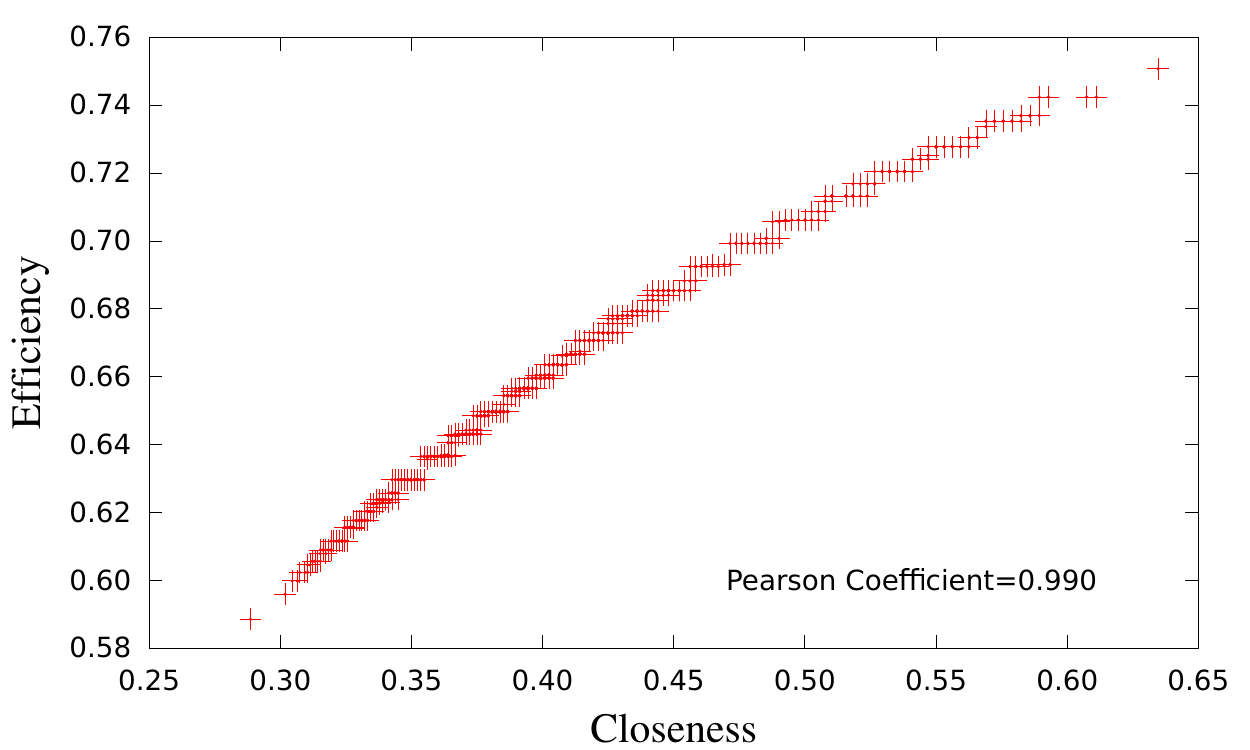}}
  \subfloat[N=500]{\includegraphics[width=0.5\textwidth]{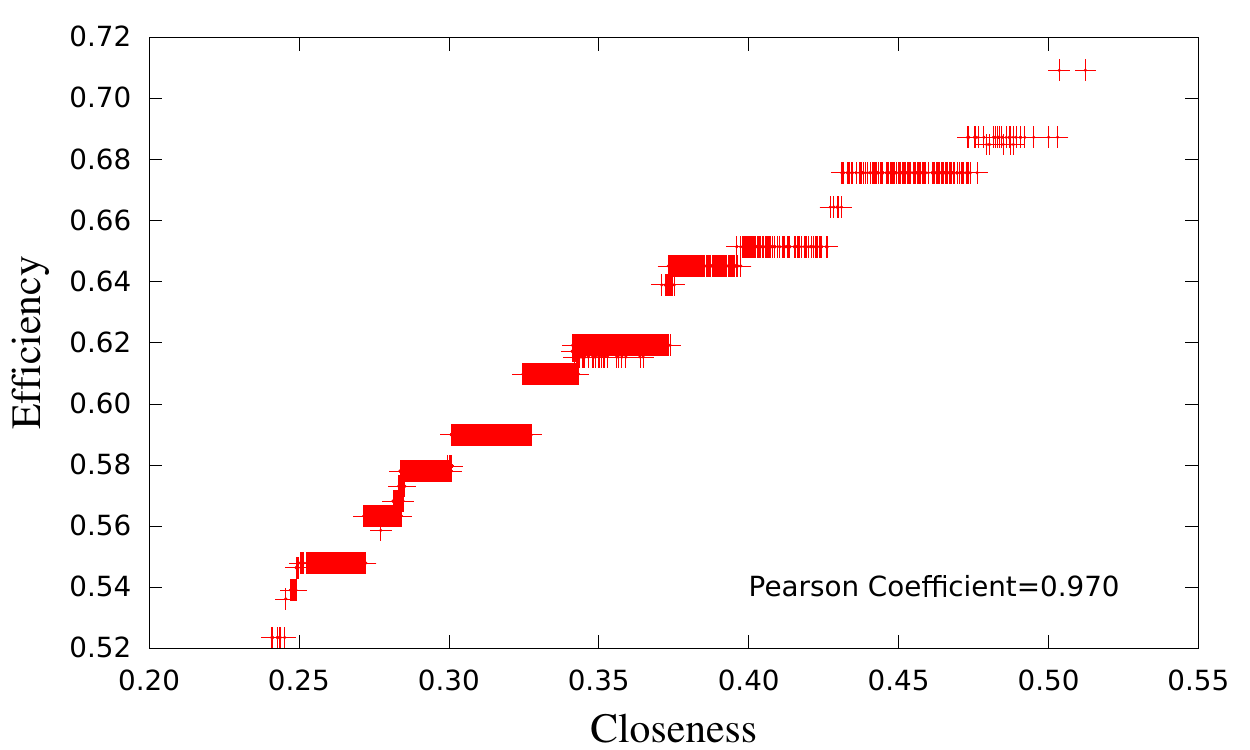}}\\
  \subfloat[N=1000]{\includegraphics[width=0.5\textwidth]{effic_dist.pdf}}
  \subfloat[N=5000]{\includegraphics[width=0.5\textwidth]{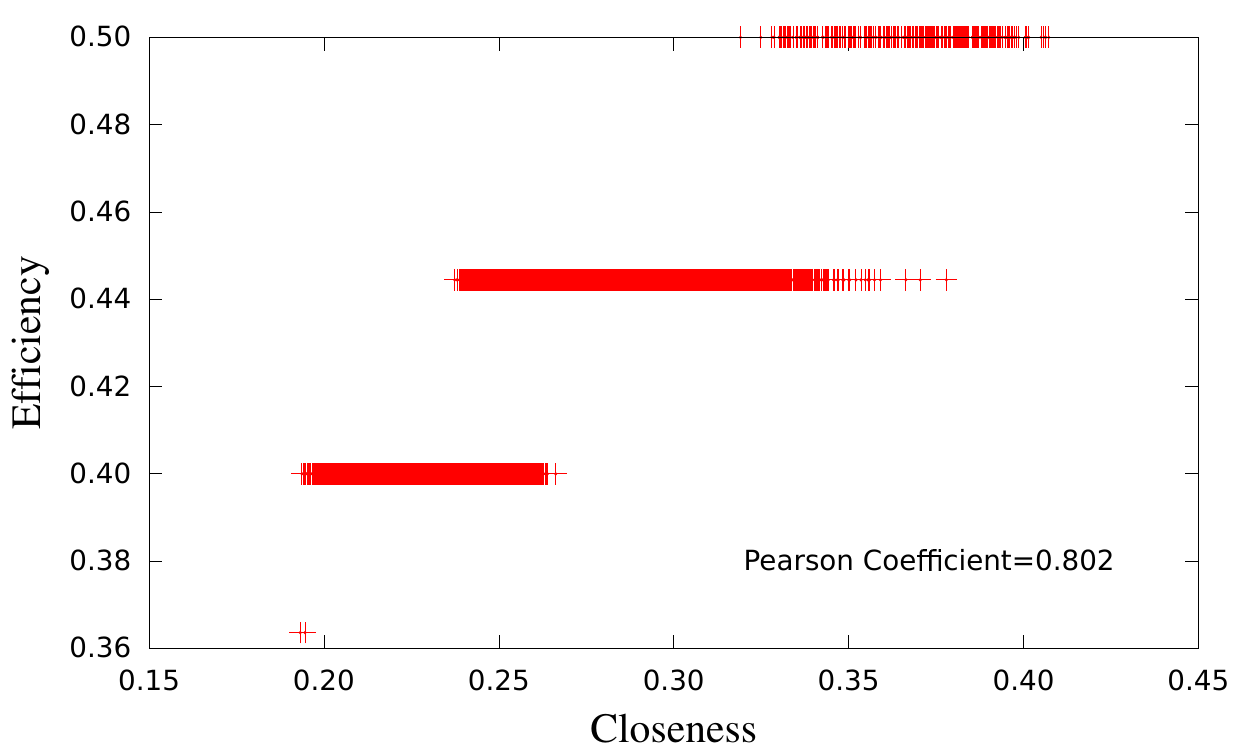}}
  \end{centering}
  \caption{Efficiency $\times$ closeness for various network sizes (BA
    model, $m=3$)}\label{effxN}
\end{figure*}

The previous experiment was performed with respect to BA networks with
fixed sizes $N=1000$.  Now we investigate the effect of $N$ on the
predictive power of closeness.  Figure~\ref{effxN} shows these
results.  The Pearson correlation coefficient decreases monotonically
with $N$, but remains high even for $N=5000$.  The quantization effect
in these results are an artifact implied because all tasks have the
same computational load, all nodes have the same computational power,
and all links transmit packets at the same rate.  With our chosen
parameter of $1$ time unit for packet transmission and $10$ time units
for task computation, nodes that are at distance~1 from the master end
their tasks at instants 13, 25, 37, 49, etc.\ while tasks from nodes
at distance~2 end at instants 16, 30, 44, 58, etc.\ and from nodes at
distance~3 at instants 19, 35, 51, 67, etc.  When the total number of
tasks is not large with respect to the number of nodes, only a few
values of execution time are be possible.

\subsection{Effect of Network Models}

\begin{figure*}[htbp]
  \begin{centering}
  \subfloat[ER $p=0.006$]{\includegraphics[width=0.5\textwidth]{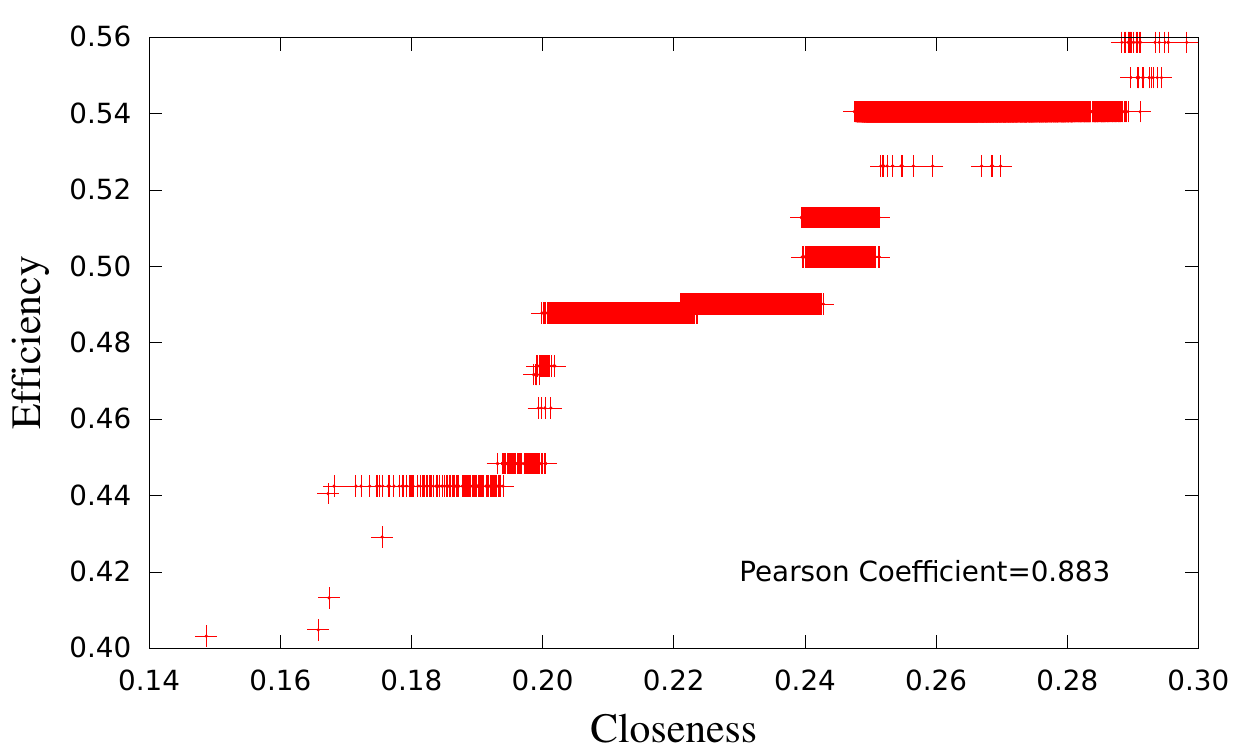}}
  \subfloat[WS $r=0.001$]{\includegraphics[width=0.5\textwidth]{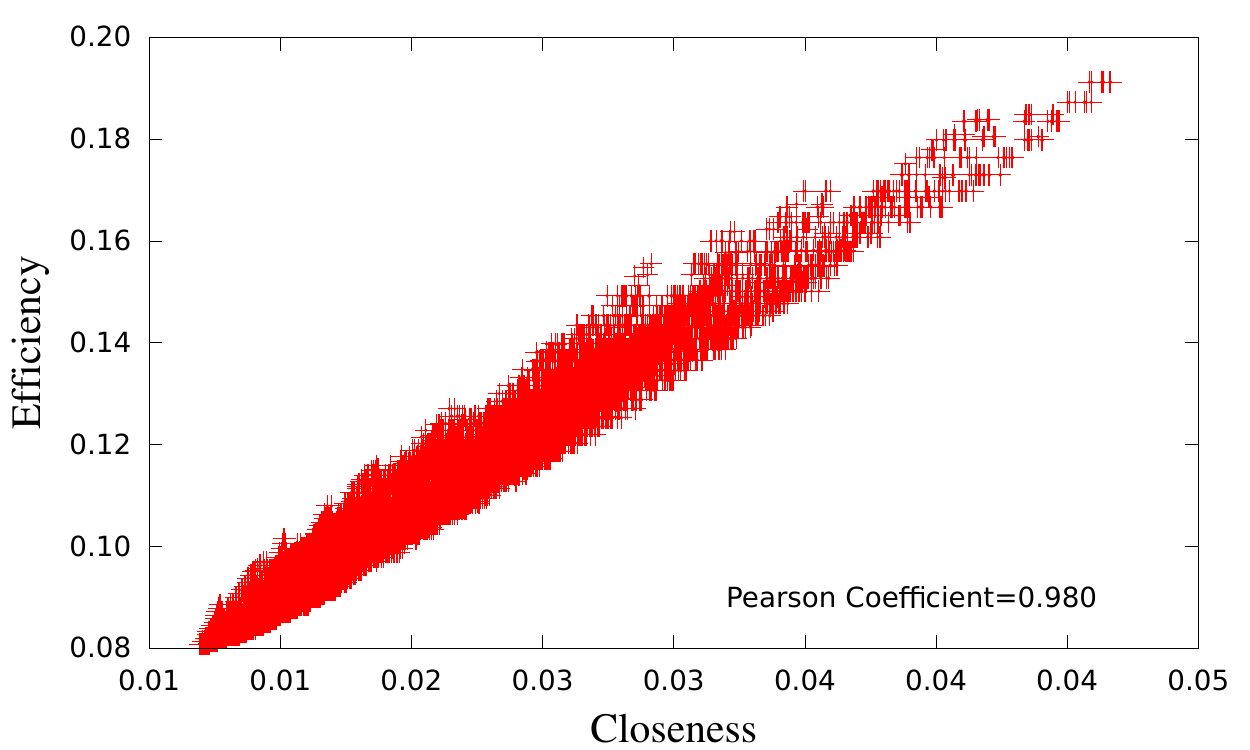}}\\
  \subfloat[WS $r=0.01$]{\includegraphics[width=0.5\textwidth]{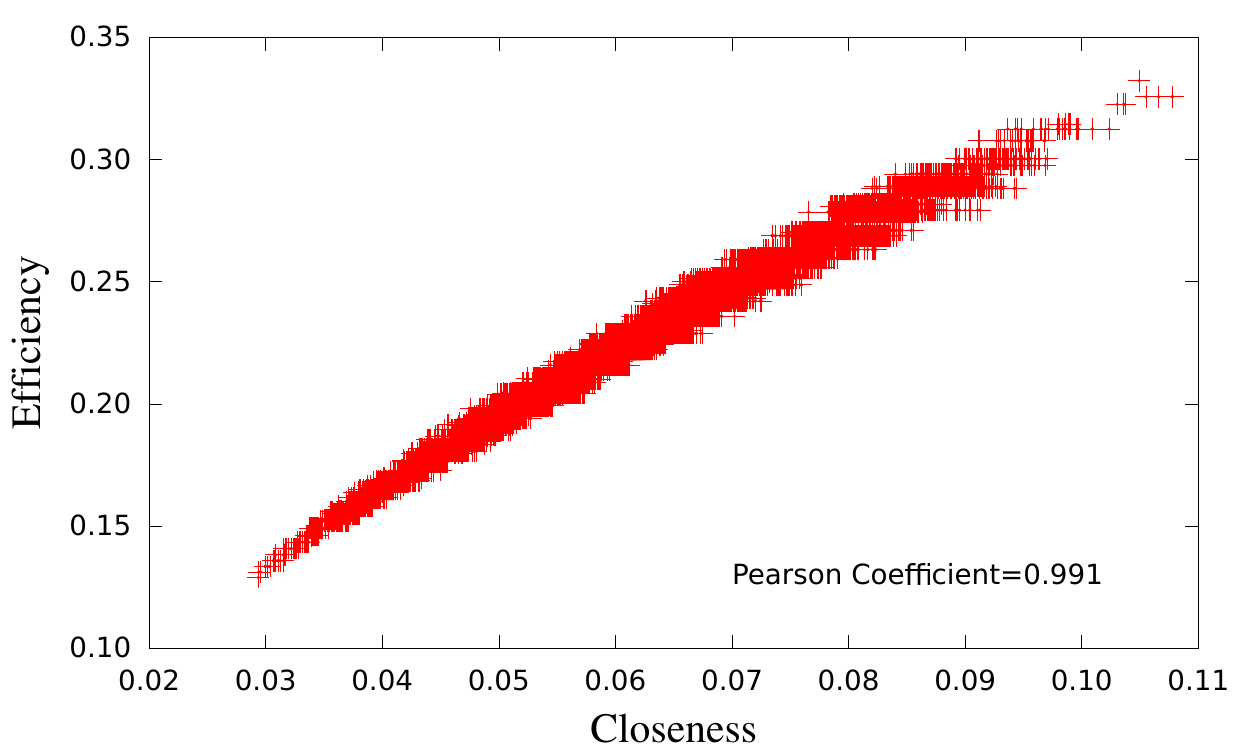}}
  \subfloat[WS $r=0.1$]{\includegraphics[width=0.5\textwidth]{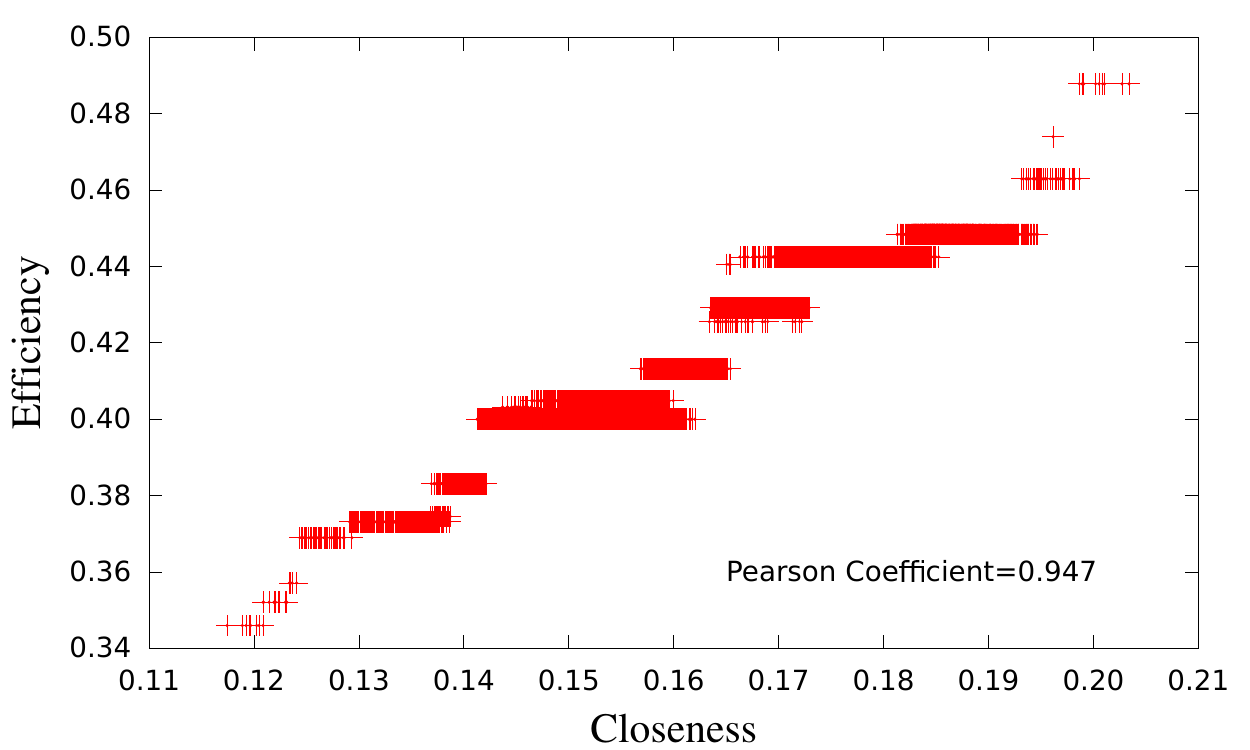}}
  \end{centering}
  \caption{Efficiency $\times$ closeness for Erdos-Renyi and
    Watts Strogatz models. The average degree is 6 for all
    cases.}\label{effxmodel}
\end{figure*}

In order to investigate the effect of network topology on the
efficiency, we performed simulations for ER and WS (for several values
of $p$) models for $N = 1000$ and $m = 3$.  The results are show in
Figure~\ref{effxmodel}.  The Pearson correlation coefficients resulted
very high for all cases.  The same quantization effect already
commented above appeared for the ER and WS with $r=0.1$.  This is a
direct consequence of the fact that these configurations have smaller
diameter.  The highest efficiency values have been observed for the ER
configuration.

\section{Conclusions}
\label{sec:conc}

The proliferation of concurrent processing systems in the last decades
as an alternative to achieving high processing speeds has emphasized
the need not only to quantify the obtained improvements, but also to
try to predict the respective performance.  Grid computing has been
receiving growing attention along the last years because of its
potential to harness available processing capabilities in a flexible
and efficient manner.  In this work, we set out to investigate how the
efficiency of grid systems can be predicted while taking into account
several topologic measurements of the topology of the grid network.
First, we quantified, through the Pearson correlation coefficient, the
potential of several measurements from complex networks to predict the
efficiency of BA networks, finding out that the closeness provided the
best alternative.  Then, we checked out the influence of the networks
size (i.e. number of nodes) on the prediction of the efficiency.
Despite a quantization effect appearing for large sizes, all tried
situations yielded very high Pearson coefficients.  Finally, we also
found that extremely good predictions can be obtained even for other
network topologies such as ER and WS.  The high values of Pearson
correlation coefficients obtained for most tested cases allow us to
choose the best master node from which to distribute the task, i.e.\
the node with the lowest average distance to other nodes would provide
the best choice for this finality.  Future works could investigate
task distribution schemes other than the master-slave configuration
addressed in the present work.

\section*{Acknowledgments}
Gonzalo Travieso is grateful to CNPq (308118/2010-3), Carlos
A. Ruggiero to FAPESP (03/08269-7), and Luciano da F. Costa to FAPESP
(2011/50761-2) and CNPq (304351/2009-1) for the financial support.

\bibliographystyle{unsrt}
\bibliography{effpred}

\end{document}